\documentclass[onecolumn, floatfix, a4paper,superscriptaddress]{revtex4}
\RequirePackage[top=12mm,bottom=12mm,left=30mm,right=30mm,head=12mm,includeheadfoot]{geometry}
\usepackage[version=3]{mhchem}
\usepackage{color}
\usepackage{braket}
\usepackage{amssymb}
\usepackage{graphicx}
\usepackage{siunitx}
\usepackage{dsfont}
\usepackage{doi}
\usepackage{verbatim}

\newcommand{\bcpo}{\ce{BiCu2PO6} }
\newcommand{\be}{\begin{equation}}
\newcommand{\ee}{\end{equation}}
\newcommand{\bs}{\begin{subequations}}
\newcommand{\es}{\end{subequations}}

\begin{document}

\author{M. Malki}
\affiliation{Lehrstuhl f\"ur Theoretische Physik I, Otto-Hahn-Str.~4, TU Dortmund, D-44221 Dortmund, Germany}
%\email{maik.malki@tu-dortmund.de}
\author{L. Splinter}
\affiliation{Lehrstuhl f\"ur Theoretische Physik I, Otto-Hahn-Str.~4, TU Dortmund, D-44221 Dortmund, Germany}
\author{G. S. Uhrig}
\affiliation{Lehrstuhl f\"ur Theoretische Physik I, Otto-Hahn-Str.~4, TU Dortmund, D-44221 Dortmund, Germany}

\title{Non-trivial topology of the quasi-one-dimensional triplons in 
the quantum antiferromagnet \bcpo }

%\maketitle

%\begin{affiliations}
% \item Lehrstuhl f\"ur Theoretische Physik 1, TU Dortmund, Germany
%\end{affiliations}

\begin{abstract}
Topological properties of physical systems are attracting tremendous interest.
Recently, magnetic solid state compounds
with and without magnetic order have become a focus. We show that \bcpo{} 
is the first gapful quantum antiferromagnet with a finite Zak phase, which characterises one-dimensional systems, and only the second
with topological non-trivial triplon excitations.
Surprisingly, in spite of the bulk-boundary correspondence
no localised edge mode occurs. This unexpected behaviour is explained
by the distinction between direct and indirect gaps among the
triplon bands. 
\end{abstract}

\maketitle

The Nobel Price 2016 awarded to Thouless, Haldane and Kosterlitz
has set an exclamation mark for the significance of topology in
physics \cite{halda17}. The research field of topology continues to expand into different areas of physics.
Topological insulators have been measured in several two and three dimensional materials \cite{konig07,hasan10,qi11,berne13,ando13,chang13}. Topological phases were also realised in a large and increasing 
variety of physical systems such as cold atoms in 
optical lattices \cite{goldm16}, photonic Floquet crystals \cite{recht13},
polaronic \cite{jacqm14},  acoustic \cite{yang15} as well as mechanical 
systems \cite{kane14,susst15}. 

Recently, quantum magnets have become a focus, in particular
magnetically ordered systems 
\cite{katsu10a,onose10,matsu11,shind13,zhang13,chisn15}.
But also a disordered valence bond crystal in a dimerised quantum magnet
has shown topologically non-trivial behaviour 
\cite{romha15,malki17a,mccla17}. Still, the number of established
compounds displaying topologically non-trivial magnetic excitations is
still extremely limited. 

The first of the two key goals of the present article
is to establish the existence of a non-trivial topological phase in 
\bcpo which represents a quasi-one-dimensional (1D)
quantum antiferromagnet \cite{plumb16,splin16,hwang16}. 
The second goal is a general one reaching far beyond
the particular material \bcpo. We show that topological
non-trivial invariants do not imply the existence 
of localised edge modes automatically. For the localisation of edge modes the
existence of an \emph{indirect} gap, i.e.\ a finite energy 
difference independent of momentum, is necessary while the 
topological phases only require the bands to be separated, i.e.\
the existence of a \emph{direct} gap at each momentum is sufficient.

Since \bcpo is essentially one-dimensional
the usual topological invariant, the Chern number 
\cite{thoul82,berry84,halda88b,hatsu93},
is not appropriate and turns out to be trivial. 
But there is another Berry phase associated to parallel transport
 in momentum space. This is the Zak phase $\Omega$ \cite{zak89} which can 
take any value between $0$ and $2\pi$ ($\Omega\in[0,2\pi)$)
because it measures the scalar product $\exp(i\Omega)=\braket{2|1}$ 
between a quantum states $\ket{1}$ at momentum 0 and the quantum state 
$\ket{2}$ taken to momentum $2\pi$ by parallel transport. 
For inversion symmetric systems the sequence
of states does not matter so that $\braket{1|2}= \braket{2|1}$ 
holds and $\Omega$ can be either $0$ or $\pi$.
Importantly, the Zak phase has been related to edge modes in strips of
graphene \cite{delpl11}. It has been measured in systems of
ultracold atoms in 1D optical lattices \cite{atala13} 
and in twisted photons \cite{carda17}.

If the eigen states as a function of a control parameter,
here a 1D momentum, can be represented in a two-dimensional plane the
system is said to possess a chiral symmetry. Then the alternative topological concept of a winding number can be used to characterise the system,
see for instance Ref.\ \cite{joshi17}.
We will show that the concept of a winding number can also be applied
to \bcpo underlining its non-trivial topological properties.

% model and material

\begin{figure}
\includegraphics[width=\textwidth]{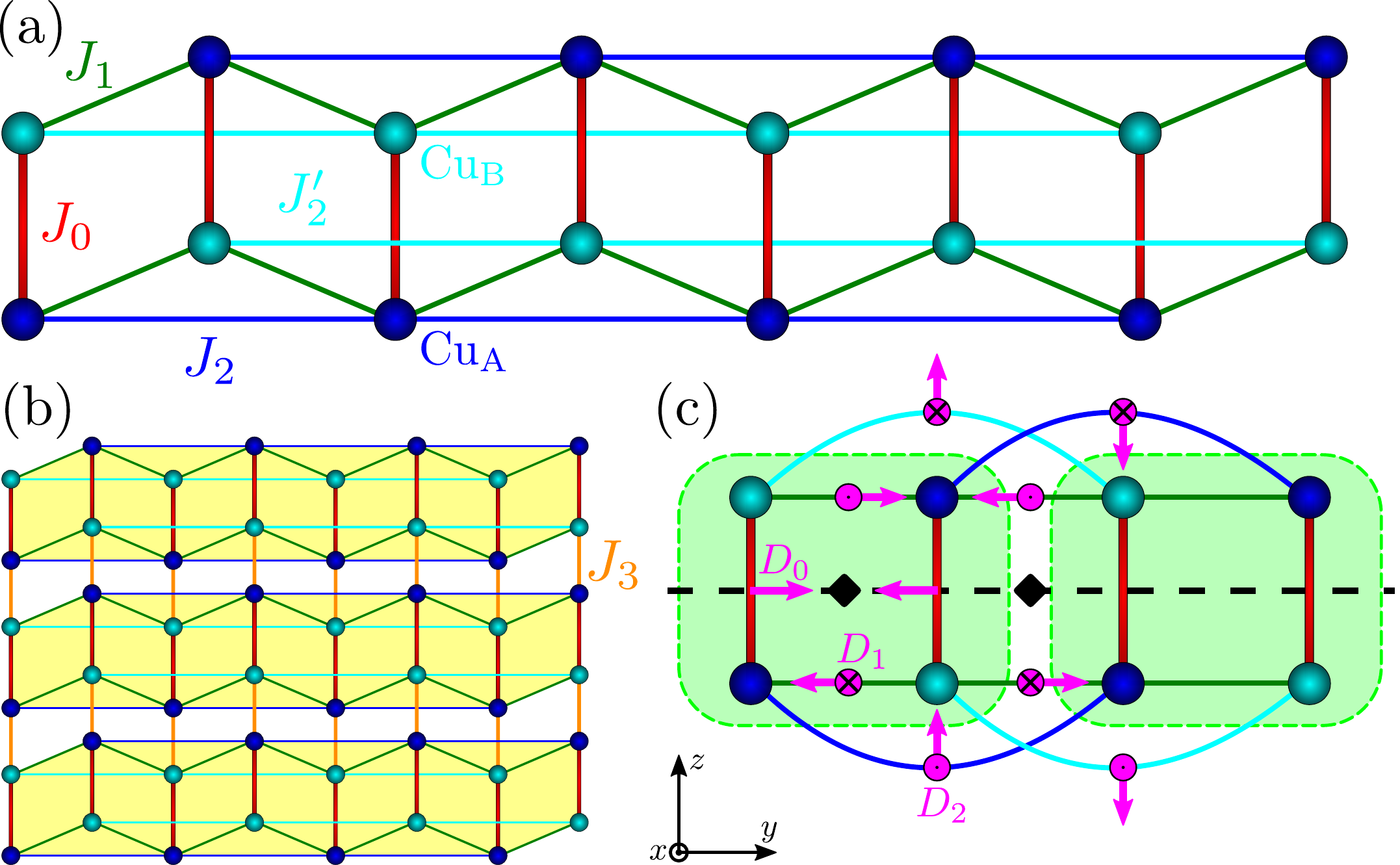}
\caption{Spin model of \bcpo. (a) The tube-like structure realises
1D frustrated spin ladders with two copper sites Cu$_A$ and
Cu$_B$. The different links stand
for different isotropic Heisenberg couplings. In case of the minimal model
the differences of copper sites are neglected so that $J_2 = J'_2$ holds. 
(b) Interladder isotropic Heisenberg coupling $J_3$ between adjacent spin ladders forming a weakly coupled two-dimensional (2D) system. 
(c) DM couplings of \bcpo. Short violet arrows display the orientation of the DM vectors in $\mathbf{D}_{ij} (\mathbf{S}_i \times 
\mathbf{S}_j)$ where we assume that the sites $i$ and $j$
are ordered with ascending $y$ or $z$ coordinate.
We highlight the inversion symmetry about the centers (black diamonds) of the plaquettes; reflection about the black dashed center line is
a symmetry of the isolated, isotropic spin ladder.}
\label{fig:lattice}
\end{figure}

The compound \bcpo is a low-dimensional quantum antiferromagnet
 with a ground state which is a valence bond solid, i.e.\ it does not
show magnetic order but a finite spin gap and a finite spin-spin correlation
length.  The spins are coupled antiferromagnetically 
in dimers which interact via further couplings 
\cite{tsirl10,plumb13,plumb16,splin16,hwang16}. 
The coupled dimers form a tube-like, frustrated spin-$1/2$ Heisenberg ladder as shown in Fig.\ 1(a). There are two types of copper ions 
Cu$_\mathrm{A}$ and Cu$_\mathrm{B}$ alternating along the ladders due to differing positions of the surrounding Bismuth ions \cite{tsirl10} (not shown
here). The 1D spin ladders form stacked 
layers with weak, but still measurable couplings
between the ladders in each layers, see Fig.\ 1(b). The couplings between layers are negligible \cite{plumb13}. The dominating couplings
are those along the spin ladders. The large atomic number ($Z = 83$) of Bismuth
induces an extraordinarily strong spin-orbit coupling (SOC) so
that the resulting magnetic exchange coupling is 
anisotropic with an important antisymmetric Dzyaloshinskii-Moriya (DM)
coupling \cite{moriy60b} and the corresponding symmetric part 
$\Gamma$ \cite{shekh92,shekh93}.

The Hamilton operator comprises isotropic Heisenberg interactions ($J_{ij}$) as well as anisotropic DM interactions ($D_{ij}^\alpha$) and symmetric anisotropic interactions ($\Gamma_{ij}^{\alpha \beta}$) given by
\begin{equation}
\mathcal{H} = \sum_{i>j} (J_{ij} \mathbf{S}_i  \cdot \mathbf{S}_j + \mathbf{D}_{ij} \cdot (\mathbf{S}_i \times \mathbf{S}_j) + \Gamma_{ij}^{\alpha \beta} S_i^\alpha S_j^\beta ) \, ,
\end{equation}
where a bold symbols represent vectors notation and $\mathbf{S}$ the spin 
vector operator. The coupling $J_0$ is the dominating rung coupling responsible 
for the dimerisation while $J_3$ describes the interladder coupling. The intraladder couplings $J_1$ and $J_2$ or $J'_2$ are the nearest neighbour and
 next-nearest neighbour couplings between the dimers.

The isotropic spin ladder is the basic building block which we describe by
dispersive triplons, i.e.\ hardcore $S=1$ quasi-particles \cite{schmi03c},
\begin{equation}
\mathcal{H}^{\mathrm{iso.} \, \mathrm{ladder}} = \sum_{k, \alpha} \omega_0(k) t_k^{\alpha, \dagger} t_k^{\alpha \phantom{, \dagger}}, 
\end{equation}
where $t_k^{\alpha, \dagger}$ creates and $t_k^{\alpha \phantom{, \dagger}}$ annihilates a triplon with momentum $k$ and 
flavour $\alpha \in \left\lbrace x, y, z \right\rbrace$ \cite{sachd90}.
The dispersion is determined systematically by continuous unitary transformations which are directly evaluated (deepCUT) in real space 
\cite{krull12}.
Fourier transformation yields the dispersion. Terms involving more than
two triplons (trilinear decay or quadrilinear interactions) are
neglected at this stage \cite{splin16}, but should be considered
on the long run \cite{chern16}.

The isotropic model leads to a degenerate triplon spectrum with six modes
at odds with experiment due to 
spin degeneracy and two dimers per unit cell. In order to include the anisotropic
terms and the interladder terms we transform in the deepCUT not only
the isotropic Hamiltonian from the spin language to the triplon language,
but also the spin operators. Then we can express the additional
anisotropic intraladder couplings and the weak interladder couplings
in terms of triplon operators. From the resulting expressions we keep again the leading bilinear terms after normal-ordering. This yields a mean-field 
description of the elementary magnetic excitations of \bcpo. 
In the isolated ladders of \bcpo, i.e.\ neglecting the
interladder coupling $J_3$, the parity with respect to reflection about the
center line is an important symmetry, see Fig.\ 1(c).
Since the creation or annihilation of
a triplon is odd the Hamiltonian can only be made up from terms with 
an even number of triplon operators \cite{schmi05b}. 

The important anisotropic couplings are responsible for lifting the degeneracy of
the triplons since they break the SU(2) spin symmetry agreeing
with experimental results \cite{plumb16,splin16,hwang16}. 
It is established that the antisymmetric DM and the symmetric $\Gamma$
coupling have to be considered together \cite{shekh92,shekh93}.
In leading order, we use
\begin{equation}
\label{eq:gamma}
\Gamma^{\alpha \beta}_{ij} = \frac{D_{ij}^\alpha D_{ij}^\beta}{2 J_{ij}} - \frac{\delta^{\alpha \beta} D_{ij}^\beta}{6 J_{ij}} ,
\end{equation} 
which results from deriving the anisotropic exchange from a Hubbard model
with SOC. The parametrisation is chosen such that $\Gamma^{\alpha \beta}_{ij}$
does not comprise an isotropic component. The isotropic components
are included in the Heisenberg couplings $J_{ij}$.

%%% Direction of DM terms
The possible directions of the DM vectors are constrained by the point group
symmetries of the lattice, see Supplementary Material. The symmetry of 
\bcpo is higher if we neglect the difference between the two copper
 sites, see Fig.\ 1, dealing with a slightly simplified model 
which we call minimal model \cite{splin16}. In this minimal model, the DM 
vectors can have components as shown in Fig.\ 1(c). Note that
 the lengths of the DM vectors is chosen arbitrarily; the goal is to 
illustrate which directions are compatible with the Moriya symmetry rules 
\cite{moriy60b}. If we take the difference between
the Cu sites into account the symmetry is reduced \cite{tsirl10}
and the possible DM vectors are given in the 
Supplementary Material. But the additionally possible
DM components are rather small because the copper sites are not
very different electronically.

The complete bilinear triplon Hamiltonian in momentum space can be 
represented in a generalised Nambu notation (up to unimportant constants)
\begin{equation}
\mathcal{H} = \frac{1}{2} \sum_{k, l} \mathbf{a}_{k, l}^\dagger 
\mathcal{M}_{k, l} \mathbf{a}_{k, l}.
\label{eq:etam}
\end{equation}
Here we combine the bosonic triplon operators into a column vector 
\begin{equation}
\mathbf{a}_{k, l} = (\mathbf{t}_{k, l}^{\top}, \mathbf{t}_{k+\pi, l}^{\top}, 
\mathbf{t}_{-k, -l}^\dagger, 
\mathbf{t}_{-k-\pi, -l}^\dagger)^{\top}
\end{equation}
with twelve components since each bold face symbol stands for three-dimensional vector
$\mathbf{t}_{k, l} = (t^x_{k, l}, t^y_{k, l}, t^z_{k, l})^{\top}$.
Hence, the Hamiltonian is described generally by a  Hermitian $12 \times 12$ matrix
\begin{equation}
\mathcal{M}_{k, l} = \begin{bmatrix}
A(k, l) & B(k, l) \\
B^\dag(k, l) & A^\top(-k, -l)
\end{bmatrix} 
\end{equation}
where the matrices $A(k, l)=A^\dag(k, l)$ and $B(k, l)=B^\top(-k, -l)$ are
 $6 \times 6 $ matrices. Note that $\mathbf{a}_{k, l}$ and thus 
$\mathcal{M}_{k, l}$ are modified relative to Ref.\ \cite{blaiz86}
in order to profit from momentum conservation.
For the inversion symmetric model
of \bcpo, further simplifications are possible, in particular for the minimal model, see Supplementary Material. The wave number $k$ corresponds to
 the direction along the ladders while the wave number $l$ corresponds to the direction perpendicular to the ladders, see Fig.\ 1(b). 

The eigen energies and eigen modes are obtained by a bosonic Bogoliubov transformation
from operators $t$ to $b$. This transformation \cite{blaiz86} is found by diagonalizing the transformed matrix 
$\widetilde{\mathcal{M}}_{k, l}:=\eta \mathcal{M}_{k, l}$ 
where  the metric $\eta$ is a diagonal matrix with components
$\eta = \mathrm{diag}(1_1, \ldots , 1_6, -1_7, \ldots , -1_{12})$. The resulting
Hamiltonain reads $\mathcal{H} = \sum_{n, k, l} \omega_{n}(k, l) b_{n, k, l}^\dagger b_{n, k, l}$ where the index $n$ labels the six different modes at given momenta $k, l$. The normal bosonic operators are given by 
\begin{align}
\label{eq:operator}
b^\dag_{n, k, l} &= \sum_{\mu=x,y,z} u_{n, k, l}^{\mu} t_{k, l}^{\mu, \dag} + 
\tilde{u}_{n, k, l}^{\mu} t_{k+\pi,l}^{ \mu, \dag} - v_{n, k,l}^{ \mu} t_{-k,-l}^{ \mu} 
- \tilde{v}_{n, k,l}^{ \mu} t_{-k-\pi,-l}^{ \mu} ,
\end{align}
where $u$ and $v$ with and without tilde are generally complex prefactors,
and its Hermitian conjugate for the annihilation operator.

The one-triplon dispersions $\omega_{n}(k,l)$ calculated in this way are used to fit the data from inelastic neutron scattering by adjusting the couplings 
($J_0$, $D_{ij}^{\alpha}$) while keeping the ratios 
$J_1/J_0 = 1.2$, $J_2/J_1 = 0.9$ and $J_3/J_0 = 0.16$ fixed
because these ratios describe the experimental wave number $k$ 
where the gap $\Delta$ occurs as well as the ratio 
between the measured lower maximum 
$\omega(k=\pi, l=2 \pi)$ and the gap $\Delta$ of the $z$-mode \cite{splin16}.
Note also that the values of the DM couplings should not be too large relative 
to the isotropic couplings in order to be realistic.

In the following, our study is based on the established minimal model for 
\bcpo \cite{splin16} assuming two identical copper ions. 
The resulting dispersions in $k$-direction in 
Fig.\ 2 agree very well with the experimental data at low energies. The discrepancies at higher energies can be explained qualitatively
by two-triplon continua implying decay processes \cite{plumb16}
which we neglect here.

\begin{figure}
\includegraphics[width=\textwidth]{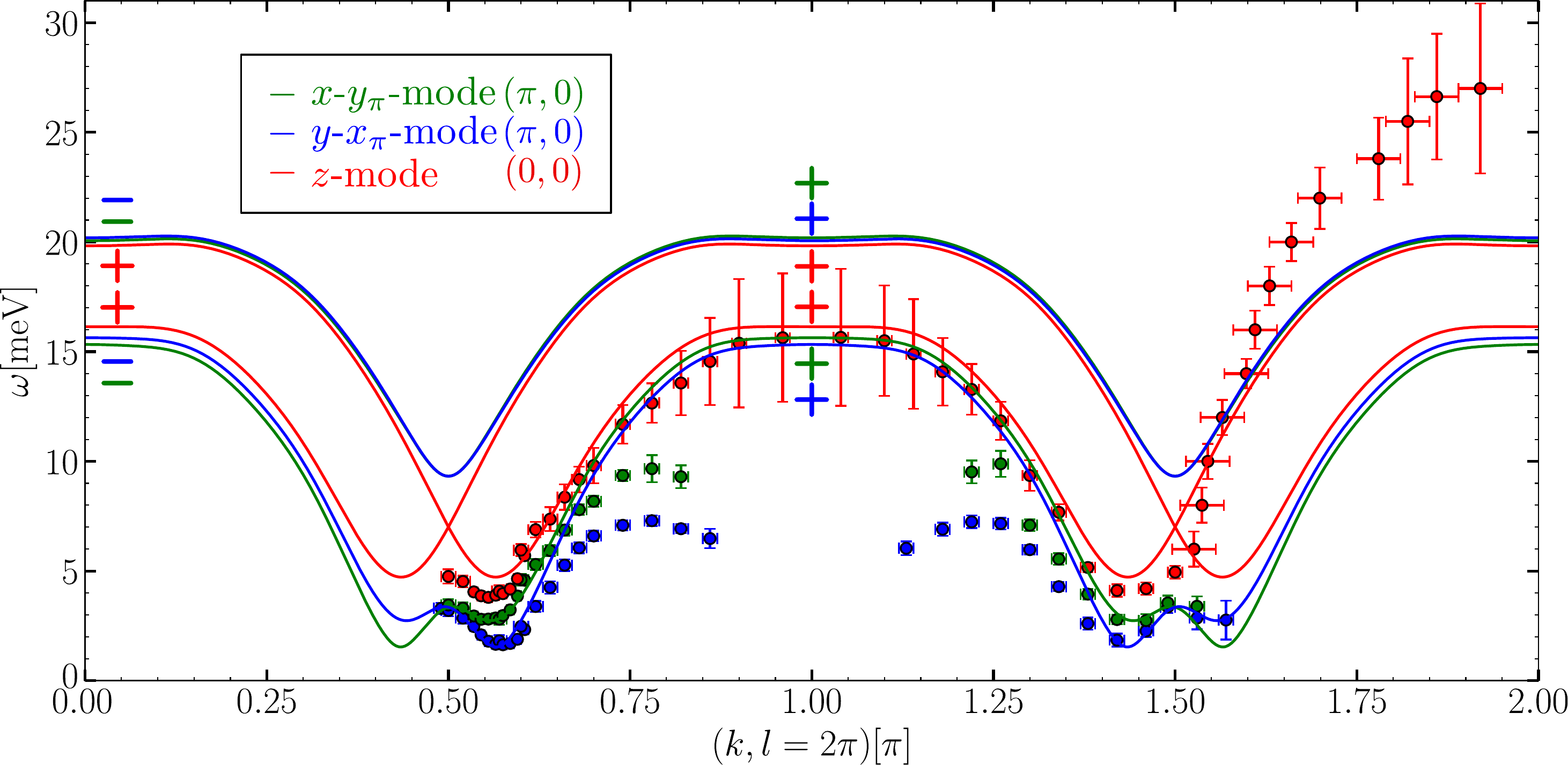} 
\caption{Computed best fit of the one-triplon dispersions for 
$J_0 = \SI{9.4}{meV}$,  $J_3=\SI{1.5}{meV}$, $J_1 = 1.2 J_0$, $J_2 = 1.09 J_0$, 
$D_1^x = 0.58 J_0$, $D_1^y = 0.73 J_0$, $D_{2, a}^z = -0.02 J_0$,
and $D_3^y = 0.02 J_0$. 
Components not listed are zero. The symbols with error bars
show the inelastic neutron scattering data from Ref.\ \cite{plumb16}.
The legend denotes the Zak phases of the modes in $k$- and $l$-direction, see main text. The plus and minus signs at the time reversal invariant wave numbers $0$ and $\pi$ indicate the parities of the modes.}
\label{fig:dispersion}
\end{figure}

For the model \eqref{eq:etam} with the appropriate fit parameters,
see Fig.\ 2, we calculate the topological properties of the triplons where we treat them as non-interacting bosons. 
This appears to be a severe approximation, 
but it is not since the deepCUT dealt with
the hardcore properties in the isotropic ladders rigorously, i.e.\ without any 
approximation. Only the weaker interladder couplings and the anisotropic couplings
are approximated by the assumption of non-interacting bosons. For low energies
and low temperatures, this is justified.

To assess the topological properties of bosonic bands one needs to generalise
the Berry curvature to bosonic systems. Even for non-interacting bosons this is 
not trivial. For fermions the scalar product of quantum states can be naturally
transferred to fermionic operators in second quantisation and the fermionic 
Bogoliubov transformations are unitary. But this does not hold for bosonic Bogoliubov transformations \cite{blaiz86} because the bosonic operators
must be normalised with respect to a symplectic product, for details see Supplementary Material. 
The operator  \eqref{eq:operator} is defined by its prefactors which 
we combine into a vector that we denote as a generalised ket state
\begin{equation}
\ket{\mathbf{k}, n}\rangle := (\mathbf{u}_{n, k, l}^\top, \tilde{\mathbf{u}}_{n, k, l}^\top, \mathbf{v}_{n, k, l}^\top, \tilde{\mathbf{v}}_{n, k, l}^\top)^\top,
\end{equation}
which is a column vector with twelve components.
The bold face symbols such as $\mathbf{u}$
stand for three-dimensional column vectors with the components 
$u^x, u^y$ and $u^z$. The symplectic product reads
\begin{subequations}
\begin{align}
& \langle\bra{\mathbf{k}_1, n_1}{\mathbf{k}_2, n_2}\rangle\rangle := \nonumber
\\
&\qquad (\mathbf{u}_{n_1, k_1, l_1}^\dag, 
\tilde{\mathbf{u}}_{n_1, k_1, l_1}^\dag, 
\mathbf{v}_{n_1, k_1, l_1}^\dag, 
\tilde{\mathbf{v}}_{n_2, k_2, l_2}^\dag) 
\ \eta \
(\mathbf{u}_{n_2, k_2, l_2}^\top, 
\tilde{\mathbf{u}}_{n_2, k_2, l_2}^\top, \mathbf{v}_{n_2, k_2, l_2}^\top, 
\tilde{\mathbf{v}}_{n_2, k_2, l_2}^\top)^\top
\\
&\qquad =\mathbf{u}_{n_1, k_1, l_1}^\dag \mathbf{u}_{n_2, k_2, l_2} +
\tilde{\mathbf{u}}_{n_1, k_1, l_1}^\dag \tilde{\mathbf{u}}_{n_2, k_2, l_2}-
\mathbf{v}_{n_1, k_1, l_1}^\dag \mathbf{v}_{n_2, k_2, l_2} -
\tilde{\mathbf{v}}_{n_1, k_1, l_1}^\dag \tilde{\mathbf{v}}_{n_2, k_2, l_2}.
\end{align}
\end{subequations}
We highlight the so far unnoted fact 
that $\widetilde{\mathcal{M}}_{k, l}$, but not $\mathcal{M}_{k, l}$,
is self-adjoint with respect to this symplectic product implying 
the well-known facts that the eigen values are real and that creation and annihilation operators
of different eigen values have to commute.

 With the above definitions, the standard relations \cite{berne13} for the Berry connection
\begin{equation}
\label{eq:vector-potential}
\mathbf{\mathcal{A}}_{n, \mathrm{sym}} (\mathbf{k}) =  \mathbf{i} \langle\bra{\mathbf{k}, n} 
{\nabla}_\mathbf{k} \ket{\mathbf{k}, n}\rangle
\end{equation}
and the Berry phase 
\begin{equation}
\label{eq:berry}
\Omega_n =  \oint \mathbf{\mathcal{A}}_n(\mathbf{k}) \mathrm{d} \mathbf{k} =  \mathbf{i} \oint \langle\bra{\mathbf{k}, n} \nabla_{\mathbf{k}} \ket{\mathbf{k}, n} \rangle \mathrm{d} \mathbf{k}
\end{equation}
can be kept.
If the closed path in the above equation encompasses the Brillouin zone, 
$\Omega_n/(2\pi)$ renders the Chern number. 
We computed the Chern number of \bcpo, but it remains trivial
even if magnetic fields are included which 
do not close the spin gap between the ground state and the lowest triplon mode. But there are other relevant topological phases in (quasi-)one-dimensional systems, notably
 the Zak phase \cite{zak89,delpl11} and the winding number \cite{joshi17}.

The Zak phase is a Berry phase computed along a closed loop in 
one direction in the Brillouin zone \cite{zak89}. Due to the periodicity in 
$k$-and $l$-space the closed loops $k\to k+2\pi$ or
$l\to l+2\pi$ allow us to define two Berry phases, 
setting the lattice constants to unity.
Each of these Berry phases can be averaged over the corresponding 
other momentum and combined into a vector $\mathbf{P}$ \cite{liu17}, which is defined by
\begin{align}
\mathbf{P} = \frac{1}{2 \pi} \int \mathbf{\mathcal{A}}_{n, \mathrm{sym}} \, \mathrm{d} k \, 
\mathrm{d} l \, .
\end{align}
The value of this vector for each triplon band is given in the
legend of Fig.\ 2. 
The $z$-mode remains topologically trivial while the coupled $x$-$y_\pi$- and 
$x_\pi$-$y$-mode display the Zak phase $(\pi,0)$, for computational 
technicalities see Supplementary Material. 

The above mentioned average does not matter in \bcpo because the
Zak phase does not depend on the wave number $l$. 
The $l$-dependence in the investigated minimal model mainly enters via the 
 isotropic term $J_3\cos(2 \pi l)$, which does not alter the eigen modes 
since this term is proportional to unity.
The small $D_3^y$ and the even smaller 
$\Gamma_{3}^{\alpha \alpha}$ barely have an impact on the dispersion and 
the eigen modes so that they do not influence the topology.
The Zak phase is constant for all values of $l$
being either zero or $\pi$. It is pinned to these
particular values in \bcpo because it is inversion symmetric, 
see Fig.\ 1(c). The transformation operator of 
inversion is given by the matrix $\Pi= 
\mathrm{diag} (1, 1, 1, -1, -1, -1, 1, 1, 1, -1, -1, -1)$ with 
$\Pi^2=\mathds{1}$ which transforms $\Pi \mathcal{M}_{k, l} \Pi 
= \mathcal{M}_{-k, -l}$ 
and hence ensures the quantisation of the Zak phase. 
We stress that the Zak phase is robust, i.e.\ small changes of the model do not
alter it. For instance, it remains the same if we pass from the minimal model
to the extended model accounting for different copper sites.
Similarly, one may reduce the values of $D_1^x$ and $D_1^y$
even by a factor 2, cf.\ Ref.\ \cite{plumb16}, 
and still retrieves the same Zak phase.
We stress, however, that they must be different $D_1^x\neq D_1^y$
to keep the Zak phase. If they are equal, the topological bands are no longer
separated so that no Zak phase can be defined or it is trivial. Note that 
$D_1^x\neq D_1^y$ is required in order to fit the experimental data. Furthermore, the
Zak phase persists in the presence of magnetic fields which do not close the spin gap 
above the ground state. This insensitivity results from the fact that the twist in the $U(1)$ 
principal fiber bundle is generated by the coupling between the $k$ and $k+\pi$ momenta. 
Thus, terms coupling at the same momentum such as the magnetic field barely destruct the Zak phase.

The momenta with $k,l\in\{0,\pi\}$ are invariant under inversion so that the
bands at these momenta have a sharply defined parities with respect to inversion
denoted by ``+'' and by ``--''  in Fig.\ 2. The products of the 
parities at $k=0$ and $k=\pi$ both at $l=2\pi$ is equal to the exponential of the Zak 
phase \cite{hughe11}. This agrees with the direct computation of the Zak phase 
in $k$-direction. This represents an alternative way to determine Zak phases.

%%% winding number
Another quantised topological index related to the Zak phase
is the winding number \cite{schny08,delpl11,li15b,joshi17}, which counts
the number of windings around a point on a 2D plane. For this concept to make
sense the Hamiltonian must have an additional symmetry, conventionally called 
chiral symmetry, so that its variation along the considered path of a control 
variable, here from $k=0$ to $k=2\pi$, can be described in a 2D plane, see 
Refs.\ \cite{delpl11,li15b,joshi17}. 
Such a chiral symmetry can be found for the minimal model of \bcpo, 
i.e.\ ignoring the difference between the copper sites, while we were not 
able to find a chiral symmetry for the extended model accounting for 
different copper sites.
The winding numbers $w$ found for the $x$-$y_\pi$- and 
$x_\pi$-$y$-mode both take the non-trivial value $w=1$, for
 details see Supplementary Material.
We emphasise, however, that the Zak phase itself is by far a more general concept
because its definition and computation does not require an additional chiral symmetry.

%%% (in)direct gap and edge states
Generically, the bulk-boundary correspondence \cite{berne13} implies that there
 must exist additional states in spatially restricted geometries of topologically non-trivial phases.
This holds if the topological invariant is quantised so that it cannot
smoothly evolve towards a trivial value on the other side of the boundary.
Edge states must exist in the gaps of strips of systems with 
finite Zak phase or finite winding number \cite{delpl11,li15b,joshi17}. 
Hence we expected this to
hold true in  \bcpo and computed the energy spectra for finite pieces
of the spin ladder pertaining to \bcpo. To our surprise we did not find
any localised edge states. We studied the states quantitatively by 
computing the inverse participation ratio (IPR) \cite{krame93} 
which is the standard measure of (non-)localised states.
If the IPR tends to zero for increasing system size the state is extended;
if it stays finite the corresponding state is localised. Here we use
the definition 
\begin{equation}
I_n = \sum_i p_{n, i}^2 = \sum_i |\langle\bra{n, i} {n, i}\rangle\rangle|^2
\end{equation}
adapted to the bosonic symplectic product and found that $I_n\to 0$ for
longer and longer spin ladders of \bcpo. 

This puzzling fact appears to be at odds with the common lore on edge modes.
But we can explain it by three arguments. First, 
the usual argument of bulk-boundary correspondence \cite{berne13}
requires the existence of states within the band gaps of the topologically
non-trivial bulk systems. But there is no argument which requires that these
states are \emph{localised}. The localisation is plausible because the 
states lie energetically within a gap and should not exist in the bulk 
far away from the boundaries. But if there is no gap the situation
is not clear a priori.

\begin{figure}
\centering
\includegraphics[width=0.5\textwidth]{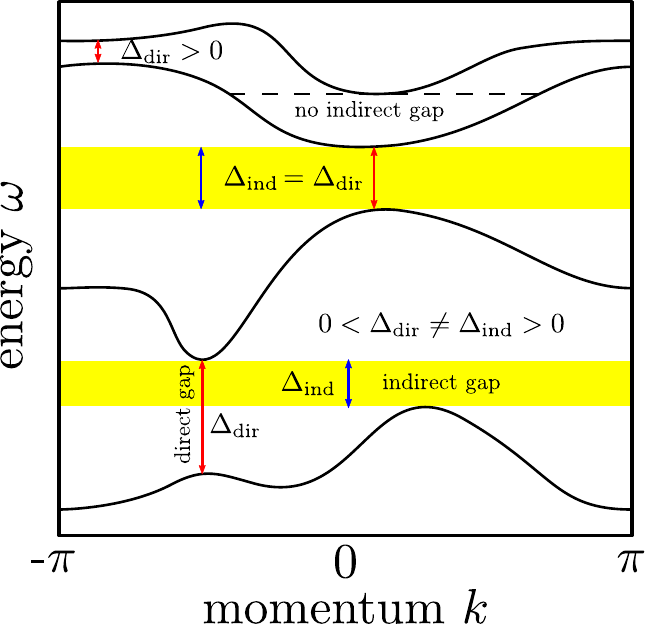} 
\caption{Illustration of a direct gap $\Delta_{\mathrm{dir}}$ at the red 
arrows and  of an indirect gap $\Delta_{\mathrm{ind}}$ (blue arrows) by the forbidden yellow area. A direct gap
is  the minimal difference between the maximal energy of a lower mode and the minimal energy of an upper mode at the \emph{same} momentum. In contrast, the indirect gap is given by a
forbidden energy interval between both modes irrespective of momentum conservation.}
\label{fig:indirect}
\end{figure}

Second, in most cases with localised edge states they lie
in an indirect gap, i.e.\ there is a whole energy interval in the bulk
without allowed states. The notion of direct and indirect gaps
is common in semiconductor physics; it is 
illustrated in Fig.\ 3. If there is an indirect gap
the gap persists even if we sum over
all momenta as one has to do in computing local densities of states.
We stress that introducing boundaries, for instance in $y$-direction, 
lifts the conservation of momentum $\hbar k$ so that generically
all these momenta hybridise. (There may be exceptions
to this hybridisation \cite{verre18a}.) 
Since the corresponding hybridising states
are extended plane waves, it is natural to expect that the resulting
states are extended as well. This is what happens in \bcpo where 
the bands are separated by direct gaps, but not by indirect gaps.

The third argument resides in the independence of the topological invariants 
in the bulk on the energies. The vector potential 
\eqref{eq:vector-potential}
and hence the Berry phase \eqref{eq:berry} depend on the eigen modes only.
They do not depend on their energies so that they
are blind to their eigen energies, i.e.\ to the dispersions. Thus, 
one can modify the Hamiltonian leaving the eigen modes completely untouched, 
but shifting their energies arbitrarily. By construction,
this does not alter the topological quantities. But it changes the system
and has an effect on the edge modes if boundaries are introduced.
To corroborate this consideration we studied the commonly
considered Su-Schrieffer-Heeger model \cite{su79}, see
Supplementary Material.
In this transparent model, we show explicitly that adding a coupling, which does
\emph{not} alter the eigen states, does alter the localisation of the edge modes.
If the indirect gap vanishes the edge modes cease to be localised, i.e.\ they
are no longer modes at the edge in the proper sense.
This finding puts the bulk-boundary correspondence generally into perspective.

To summarise, we analysed the available inelastic neutron scattering 
data in the framework of a magnetic valence bond crystal with triplons
as elementary excitations. Within the resulting model, we computed the 
Zak phase as generic one-dimensional topological invariant; it takes 
the non-trivial value $\pi$. Due to
inversion symmetry it has to be quantised in multiples of $\pi$.
The non-trivial value is robust against not too large changes of the 
DM couplings. They may even vary by a factor of two, but it is important
that $D_1^x\neq D_1^y$ holds.
The topological character of \bcpo is supported by the winding number
$w=-1$ based on the chiral symmetry of the minimal model.

Remarkably, we found that in spite of the topological bulk
properties no \emph{localised} edge modes occur. We clarified 
this unexpected finding by the distinction of direct and indirect gaps.
Only the existence of an indirect gap warrants the localisation
of edge modes. We point out that the standard bulk-boundary
correspondence implies the existence of modes within the gaps
separating the topological non-trivial bands, \emph{but} it does not
imply localisation. This has been corroborated by a comprehensive study
of the paradigmatic Su-Schrieffer-Heeger model. 

Our results identify \bcpo as the first 
disordered quantum antiferromagnet with finite quantized Zak phase
and the second disordered antiferromagnet with topologically non-trivial 
eigen modes. So far, only \ce{SrCu2(BO3)2} had been known for its non-trivial triplon excitations. Further search for low-dimensional disordered quantum
magnets with finite Zak phases or finite Chern numbers is to be expected.
On the conceptual level, the scenario of delocalisation of edge modes
deserves further investigation in all conceivable physical realisations.

%\begin{methods}
%method
%\end{methods}

%% Put the bibliography here, most people will use BiBTeX in
%% which case the environment below should be replaced with
%% the \bibliography{} command.

% \begin{thebibliography}{1}
% \bibitem{dummy} Articles are restricted to 50 references, Letters
% to 30.
% \bibitem{dummyb} No compound references -- only one source per
% reference.
% \end{thebibliography}

%\bibliographystyle{SciPost_bibstyle}
%\bibliography{liter10}

%% Here is the endmatter stuff: Supplementary Info, etc.
%% Use \item's to separate, default label is "Acknowledgements"

%\begin{addendum}
\begin{itemize}
 \item We acknowledge useful discussions with Christoph H. Redder and 
Joachim Stolze and provision of the experimental data
by Kemp Plumb and Young-June Kim. Financial support (MM) was 
given by the Studienstiftung des Deutschen Volkes and  by the Deutsche Forschungsgemeinschaft 
and the Russian Foundation of Basic Research through the transregio TRR 160.
 \item %[Competing Interests] 
Competing Interests:
The authors declare that they have no
competing financial interests.
 \item %[Correspondence] 
Correspondence:
Correspondence and requests for materials
should be addressed to M.M.~(email: maik.malki@tu-dortmund.de).
%\end{addendum}
\end{itemize}

%%%%%%%%%% Merge with supplemental materials %%%%%%%%%%
\pagebreak
%\widetext
%\begin{center}
%\textbf{\large Supplementary Materials}
%\end{center}
%%%%%%%%%% Merge with supplemental materials %%%%%%%%%%
%%%%%%%%%% Prefix a "S" to all equations, figures, tables and reset the counter %%%%%%%%%%
\setcounter{equation}{0}
\setcounter{figure}{0}
\setcounter{table}{0}
\setcounter{page}{1}
\makeatletter
\renewcommand{\theequation}{S\arabic{equation}}
\renewcommand{\thefigure}{S\arabic{figure}}
\renewcommand{\bibnumfmt}[1]{[S#1]}
\renewcommand{\citenumfont}[1]{S#1}
%%%%%%%%%% Prefix a "S" to all equations, figures, tables and reset the counter %%%%%%%%%%

\hspace{4cm}

{\noindent \textbf{\large Supplementary Note 1: Symmetry analysis of \bcpo}}\\

\noindent The direction of the Dzyaloshinskii-Moriya (DM) vectors $\mathbf{D}_m,
m\in\{0,1,2,3\}$ are restricted due to the symmetries of the system. These restrictions are formulated by the five selection rules of 
Moriya \cite{s_moriy60b} which relate the different couplings based on the point
 group symmetries of the system. For the sake of completeness, we present these five selection rules here briefly. 
Moriya established them by considering two interacting ions with spins whose positions we label with $A$ and $B$. 
The center of the connecting line $\overline{AB}$ is denoted by $C$. 
\begin{itemize}
\item[1$^\text{st}$]{If $C$ presents a center of inversion, then $\mathbf{D}=0$ holds.}
\item[2$^\text{nd}$]{If there is a mirror plane  perpendicular to $\overline{AB}$ and passing through $C$, then $\mathbf{D} \perp \overline{AB}$ is valid.}
\item[3$^\text{rd}$]{If a mirror plane including the positions $A$ and $B$ is present, the vector $\mathbf{D}$ is perpendicular to this mirror plane.}
\item[4$\text{th}$]{In case of a two-fold rotation axis perpendicular to the line 
$\overline{AB}$ and passing through $C$, then $\mathbf{D}$ is
perpendicular to this two-fold rotation axis.}
\item[5$^\text{th}$]{If there is an $n$-fold axis ($n \geq 2$) passing along
 $\overline{AB}$, the relation $\mathbf{D} \parallel \overline{AB}$ is valid.}
\end{itemize}

Besides the information that specific $\mathbf{D}_{ij}$ components are forbidden due to point group symmetries of the single bonds
one can additionally obtain information on the signs of the possible
$\mathbf{D}_{ij}$ along the ladder by considering translations and glide
reflections. Likewise the parity of the components
relative to reflection about the center line, see Fig.\ 1(c) in the main 
article, can be elucidated. 
This parity determines whether a term contributes to the dispersions 
on the level of bilinear Hamiltonians or not \cite{s_schmi05b}. 

\begin{figure}[htb]
\centering
\includegraphics[width=0.6\textwidth]{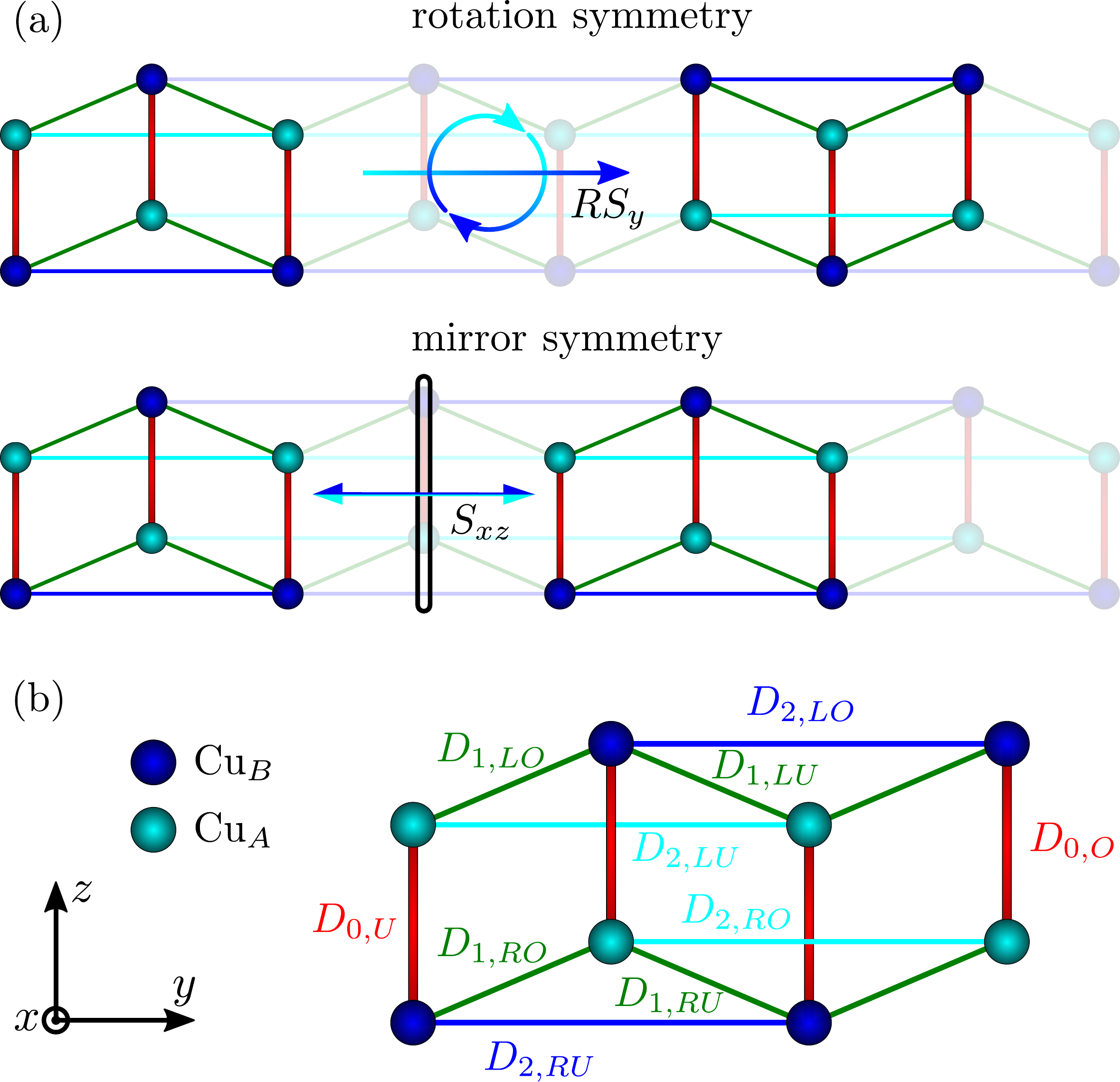} 
\caption{Symmetry analysis. (a) Illustration of the two symmetry operations 
$RS_y$ and $S_{xz}$ leaving the model of \bcpo invariant. (b) 
Notation of the various $\mathbf{D}$ vectors in \bcpo.}
\label{fig:sym}
\end{figure}

If we neglect the difference between the two copper ions \ce{Cu_A} and 
\ce{Cu_B} 
we arrive at the minimal model of \bcpo with the possible DM components
shown in Fig.\ 1(c) of the main article.
Taking into account the difference between the two copper sites \cite{s_tsirl10} the symmetry of the lattice is lower so that more 
$\mathbf{D}_{ij}$ components are allowed. Then only the following two 
symmetries of the crystal structure are present:
\begin{enumerate}
\item $RS_y$: Rotation by $\pi$ around the $y$-axis located in the middle of
the spin ladder and a shift by half a unit cell.
\item $S_{xz}$: Reflection at the $xz$-plane located at a dimer.
\end{enumerate}
These two symmetry operations are shown in Fig.\ \ref{fig:sym}(a). 
In Fig.\ \ref{fig:sym}(b) the notation of the various DM vectors is shown.

The determined symmetries imply the following constraints. The vector 
$\mathbf{D}_0$ only has a $y$-component due to the third selection rule 
based on the symmetry $S_{xz}$. The $RS_y$ symmetry  yields the relation
\begin{align}
RS_y(\mathbf{D}_{0, U}) = - \mathbf{D}_{0, O}.
\label{eq:D0y}
\end{align}
After a $RS_y$ rotation the stipulated sequence of the spin operators within the 
term $\mathbf{D}_{ij} (\mathbf{S}_i \times \mathbf{S}_j)$ (according to ascending $y$- and $z$-coordinate) must be 
recovered by swapping the spin operators. Thus, Eq.\ \eqref{eq:D0y} shows the alternating behaviour of $D_0^y$ along the legs. The symmetry analysis of the 
$\mathbf{D}_1$ bond leads to the relations
\begin{subequations}
\begin{align}
S_{xz} (\mathbf{D}_{1, LU}) &= \mathbf{D}_{1, LO} \label{eq:D1sx}\\
S_{xz} (\mathbf{D}_{1, LO}) &= \mathbf{D}_{1, LU} \\
S_{xz} (\mathbf{D}_{1, RU}) &= \mathbf{D}_{1, RO} \\
S_{xz} (\mathbf{D}_{1, RO}) &= \mathbf{D}_{1, RU} \\
RS_{y} (\mathbf{D}_{1, LU}) &= \mathbf{D}_{1, RO} \label{eq:D1rsy}\\
RS_{y} (\mathbf{D}_{1, LO}) &= \mathbf{D}_{1, RU} \\
RS_{y} (\mathbf{D}_{1, RU}) &= \mathbf{D}_{1, LO} \\
RS_{y} (\mathbf{D}_{1, RO}) &= \mathbf{D}_{1, LU} .
\end{align}
\end{subequations}

To clarify the properties of $\mathbf{D}_1$ we start with an arbitrary vector
\begin{align}
\mathbf{D}_{1, LU} =  c_x \mathbf{e}_x + c_y \mathbf{e}_y + c_z \mathbf{e}_z
\end{align}
where $\mathbf{e}_\mu$ are unit vectors in the directions
indicated by the subscript and $c_\mu$ are real coefficients. Applying Eqs.\ \eqref{eq:D1sx} and \eqref{eq:D1rsy} to this ansatz 
for $\mathbf{D}_{1, LU}$ we obtain
\bs
\begin{align}
\mathbf{D}_{1, LO} &= \phantom{-} c_x \mathbf{e}_x - c_y \mathbf{e}_y + c_z \mathbf{e}_z 
\\
\mathbf{D}_{1, RO} &= - c_x \mathbf{e}_x + c_y \mathbf{e}_y - c_z \mathbf{e}_z.
\end{align}
\es
The first condition determines that the $x$- and $z$-component are uniform 
while the $y$-component is alternating along the ladder. The second condition
 indicates that all three components have odd parity since the translation to 
$\mathbf{D}_{1, RU}$ changes the sign of the $y$-component as well
 so that all coefficients acquire a negative sign. 

In the same way, we 
investigate $\mathbf{D}_2$. 
Applying both symmetry operations to $\mathbf{D}_2$ yields
\begin{subequations}
\begin{align}
S_{xz} (\mathbf{D}_{2, LU}) &= \mathbf{D}_{2, LU} \label{eq:D2sx}\\
S_{xz} (\mathbf{D}_{2, LO}) &= \mathbf{D}_{2, LO} \\
S_{xz} (\mathbf{D}_{2, RU}) &= \mathbf{D}_{2, RU} \\
S_{xz} (\mathbf{D}_{2, RO}) &= \mathbf{D}_{2, RO} \\
RS_{y} (\mathbf{D}_{2, LU}) &= \mathbf{D}_{2, RO} \label{eq:D2rsy}\\
RS_{y} (\mathbf{D}_{2, LO}) &= \mathbf{D}_{2, RU} \\
RS_{y} (\mathbf{D}_{2, RU}) &= \mathbf{D}_{2, LO} \\
RS_{y} (\mathbf{D}_{2, RO}) &= \mathbf{D}_{2, LU} \, .
\end{align}
\end{subequations}
Again, we start from the general ansatz
\begin{align}
\mathbf{D}_{2, LU} =  d_x \mathbf{e}_x + d_y \mathbf{e}_y + d_z \mathbf{e}_z \, .
\end{align}
Using Eq.\ \eqref{eq:D2sx} we easily  see that the $y$-component has to vanish. In contrast, using Eq.\ \eqref{eq:D2rsy} does not lead to an 
unambiguous solution because we obtain
\begin{align}
\mathbf{D}_{2, RO} =  - d_x \mathbf{e}_x - d_z \mathbf{e}_z \, .
\end{align}
Each component can fulfil this condition in two different ways. Either the component is alternating along the ladder with even parity or it is uniform 
along the ladder with odd parity. Thus, the $\mathbf{D}_2$-vector is 
generally expressed by the superposition of both possibilities
\begin{subequations}
\label{eq:D2}
\begin{align}
D_2^x &= D_{2, a}^x + D_{2, u}^x \\
D_2^z &= D_{2, a}^z + D_{2, u}^z 
\end{align}
\end{subequations}
where subscript $a$ stands for ``alternating'' and $u$ for ``uniform''.

Considering the fact that the differences between the copper ions are small 
\cite{s_tsirl10} we may neglect them altogether  which allows us to conclude 
\cite{s_splin16}  $D_2^x = D_{2, u}^x$ and $D_2^z = D_{2, a}^z$. 
Thus, we conclude that the uniform $x$-component and the alternating
$z$-component predominate. Arbitrary components as in Eq.\ \eqref{eq:D2} are allowed, but decisive contributions only come from the alternating even parity 
$z$-component and the uniform odd parity $x$-component. 

Note that we neglect potential differences between  $\mathbf{D}_2$ on the
$J_2$ bond and $\mathbf{D}'_2$ on the $J_2'$ bond because they have odd
parity and do not contribute on the bilinear level anyway. The potential differences in the ensuing symmetric $\Gamma$-terms are neglected as well
because of their barely measurable impact. 

The results of the symmetry analysis are collected 
in Tab.\ \ref{tab:symmetry}. Since the $\Gamma$-couplings result
from the $D$-couplings according to Eq.\ (3) in the main article
one can establish a similar table for the $\Gamma$-components based on Tab.\
 \ref{tab:symmetry}. The property of being alternating/odd corresponds to a minus sign while uniform/even to a plus sign in the DM components. Thus by 
multiplying $\pm 1$ to the DM components in Eq.\ (3) one arrives at
 the resulting properties of the $\Gamma$-components. 

Finally, we remark that the orientation of the $\mathbf{D}_3$-vector, which belongs to the interladder coupling, is analogous to the
$\mathbf{D}_0$-vector. The $\mathbf{D}_3$-vector couples two adjacent ladders contributing to the transversal dispersion. No  parity can be defined because
the reflection about the center line refers to a symmetry within each ladder separately.

\begin{table}
\centering
\caption{Behaviour of the sign of the $D$-components along the legs of the spin ladder and their parity with respect to the symmetry $S_{xy}$
(reflection with respect to the center line of the spin ladder, see
Fig.\ 1(c) of the main text).  $D$-components which are not listed are 
forbidden due to the selection rules of Moriya \cite{s_moriy60b}.}
\label{tab:symmetry}
\medskip
\begin{tabular}{ccc}
\hline
couplings & along the legs & parity\\
\hline
$D_0^y$ & alternating & odd\\
$D_1^x$ & uniform & odd\\
$D_1^y$ & alternating & odd\\
$D_1^z$ & uniform & odd\\
$D_{2, a}^x$ & alternating & even\\
$D_{2, u}^x$ & uniform & odd\\
$D_{2, a}^z$ & alternating & even\\
$D_{2, u}^z$ & uniform & odd\\
$D_3^y$ & alternating & N/A \\
\hline
\end{tabular}
\end{table}

\bigskip

{\noindent \textbf{\large Supplementary Note 2: 
Matrix representation of the bilinear Hamilton operator}}\\

\noindent The general expression in Nambu representation of the complete bilinear Hamiltonian in quasi-momentum space is given up to unimportant 
constants by 
\begin{equation}
\label{eq:hamiltonian12}
\mathcal{H} = \frac{1}{2} \sum_{k, l} \mathbf{a}_{k, l}^\dagger 
\mathcal{M}_{k, l} \mathbf{a}_{k, l} \quad \mathrm{with} 
\quad \mathcal{M}_{k, l} = \begin{bmatrix}
A(k, l) & B(k, l) \\
B^\dagger(k, l) & A^{\top}(-k, -l)
\end{bmatrix} = \mathcal{M}^\dag_{k, l} 
\end{equation}
and the twelve-dimensional Nambu spinor
$\mathbf{a}_{k, l} = (\mathbf{t}_{k, l}^{\top}, \mathbf{t}_{k+\pi, l}^{\top}, 
\mathbf{t}_{-k, -l}^\dagger, \mathbf{t}_{-k-\pi, -l}^\dagger)^{\top}$,
see Eq.\ (5) in the main text, using 
$\mathbf{t}_{k, l} = (t^x_{k, l}, t^y_{k, l}, t^z_{k, l})^{\top}$. 
Note that the sum in \eqref{eq:hamiltonian12} runs over all values of 
$l\in[0,2\pi)$ (lattice constant set to unity) 
in the Brillouin zone while it runs only over the values
$k\in[0,\pi)$, i.e.\ over half the Brillouin zone. The reason is that the above
Nambu spinor addresses $k$ and $k+\pi$ simultaneously.

The $12\times12$ matrix $\mathcal{M}_{k, l}$ is composed of the two  
$6 \times6$ matrices $A$ and $B$ which are again made up by $3\times 3$ matrices
\begin{subequations} 
\begin{align}
A(k, l) &= 
\begin{pmatrix}
A_1(k) + B_1(k, l) & B_2(k, l) \\
B_2^\dag(k, l) & A_1(k + \pi) + B_1(k + \pi, l)
\end{pmatrix} \\
B(k, l) &= 
\begin{pmatrix}
B_1(k, l) & B_2(k, l) \\
B_2^\dag(k, l) & B_1(k + \pi, l) \, .
\end{pmatrix}
\end{align}
\end{subequations} 
The $3 \times 3$ matrices are derived to be
\begin{subequations} 
\begin{align}
A_1(k) &= 
\begin{pmatrix}
\omega_0(k) & \mathrm{i} h_z & -\mathrm{i} h_y\\
-\mathrm{i} h_z & \omega_0(k) & \mathrm{i} h_x\\
\mathrm{i} h_y & -\mathrm{i} h_x & \omega_0(k) \\
\end{pmatrix} \\
B_1(k, l) &= 
\begin{pmatrix}
F^x(k, l)& 0 & \Gamma_1^{xz}(k) + \Gamma_2^{xz}(k)\\
0 & F^y(k, l) & 0\\
\Gamma_1^{xz}(k) + \Gamma_2^{xz}(k) & 0 & F^z(k, l) \\
\end{pmatrix} \\
B_2(k, l) &= 
\begin{pmatrix}
0 & - \mathrm{i} (\Gamma_1^{xy}(k) - D_2^z(k)) & -\mathrm{i} D_3^y (k, l)\\
- \mathrm{i} (\Gamma_1^{xy}(k) + D_2^z(k)) & 0 & - \mathrm{i} (\Gamma_1^{yz}(k) - D_2^x(k)) \\
\mathrm{i} D_3^y (k, l) & - \mathrm{i} (\Gamma_1^{yz}(k) +  D_2^x(k)) & 0
\end{pmatrix} \, .
\end{align}
\end{subequations}

\begin{table}
\centering
\caption{The coefficients $\omega_\delta$ in order to describe the 
dispersion of the isotropic spin ladder as well as the prefactors $a_\delta$ 
to transform the spin operators are calculated by using the deepCUT method
for the ratios $J_1/J_0=1.2$ and $J_2/J_1=0.9$. The values for higher $\delta$
are small enough to be neglected.}
\label{tab:CUT}
\medskip
\begin{tabular}{lcc}
\hline
$\delta$ & \centering $\omega_\delta$ & $a_\delta$\\
\hline
0 & \phantom{-}1.5499384208488\phantom{00000} & \phantom{-}0.3874491109155713\phantom{00} \\
 1 & \phantom{-}0.358817770492231\phantom{000} &-0.05165001704799924\phantom{0} \\
 2 &\phantom{-}0.524739087510573\phantom{000} &-0.08095884805094124\phantom{0} \\
3 &-0.209722209664048\phantom{000} &\phantom{-}0.03713614889687351\phantom{0} \\
 4 &-0.160344853773972\phantom{000} &\phantom{-}0.0219291397751164\phantom{00} \\
 5 &\phantom{-}0.0967516245738429\phantom{00} &-0.01719462494862808\phantom{0} \\
6 &\phantom{-}0.010462389004026\phantom{000} &-0.004727305201296136 \\
 7 &-0.0347043572019398\phantom{00} &\phantom{-}0.01024208259455439\phantom{0} \\
 8 &\phantom{-}0.000112462598212057 &-0.001628782296091526 \\
 9 &\phantom{-}0.0139297388647789\phantom{00} &-0.00497492501969249\phantom{0} \\
 10 &-0.00637707478352971\phantom{0} &\phantom{-}0.002315960919757644 \\
 11 &-0.00403742286524941\phantom{0} &\phantom{-}0.001621270078823474 \\
 12 &\phantom{-}0.00429559542625067\phantom{0} &-0.001835116321222724 \\
 13 &\phantom{-}0.000461321168694168 & \\
\hline
\end{tabular}
\end{table}

The dispersion of the isotropic spin ladder is calculated by deepCUT method 
\cite{s_splin16,s_krull12} yielding
\begin{equation}
\omega_0(k) = \sum_{\delta = 0}^{13} \omega_\delta \cos(\delta k) \, .
\end{equation}
The coefficients  $\omega_\delta$ are given in Tab.\ \ref{tab:CUT}. 
Similarly, the transformation of the spin operators to triplon operators
\begin{equation}
S^\mu_{0,L} = -S^\mu_{0,R} = \sum_{\delta=-12}^{12} a_{|\delta|} 
(t^\mu_\delta+ t^{\mu,\dag}_\delta) + \text{bilinear and higher terms}
\end{equation}
yields the amplitudes $a_\delta$ also given in Tab.\ \ref{tab:CUT}.
The spin operators are labelled with subscript left ($L$) and right ($R$)
spin in a dimer referring to the two legs of each ladder.
Bilinear or higher products of triplon operators are neglected in our
approach to the transformation of the spin operator. The Fourier transform
\begin{equation}
a(k):=\sum_{\delta = -12}^{12} a_{|\delta|} \cos(\delta k) 
\end{equation}
yields the momentum dependent amplitude $a(k)$ which appears generically
in effective triplon Hamiltonians \cite{s_uhrig04,s_uhrig05a,s_splin16}.
The Hamiltonian also includes a general uniform magnetic field 
$\mathbf{h} = (h_x, h_y, h_z)^\top$ given by 
$\mathcal{H}_\text{Zeeman}= - \mathbf{h} \sum_i \mathbf{S}_i$.

Further variables introduced for clarity are
\begin{align}
F^\mu(k, l) = d(k, l) + \Gamma_0^{\mu \mu}(k) + \Gamma_1^{\mu \mu}(k) + 
\Gamma_2^{\mu \mu}(k) + \Gamma_3^{\mu \mu}(k, l) \quad \mathrm{with} \quad 
\mu \in \left\lbrace x, y, z \right\rbrace 
\end{align}
and
\begin{subequations}
\begin{align}
d(k, l) &= - 2 J_3 \cos(2\pi l) a^2(k) \\
\Gamma_0^{\mu \mu}(k) &=  -2 \Gamma_0^{\mu \mu} a^2(k) \\
\Gamma_1^{\mu \mu}(k) &= \phantom{-} 4 \Gamma_1^{\mu \mu} a^2(k) \cos(k)\\
\Gamma_2^{\mu \mu}(k) &= \phantom{-} 4 \Gamma_2^{\mu \mu} a^2(k) \cos(2 k) \\
\Gamma_3^{\mu \mu}(k, l) &=  - 2 \Gamma_3^{\mu \mu} a^2(k) \cos(2 \pi l) \\
\Gamma_1^{xy}(k) &= \phantom{-} 4 \Gamma_1^{xy} a(k) a(k+\pi) \sin(k) \\
\Gamma_1^{yz}(k) &= \phantom{-} 4 \Gamma_1^{yz} a(k) a(k+\pi) \sin(k) \\
\Gamma_1^{xz}(k) &= \phantom{-} 4 \Gamma_1^{xz} a(k)^2 \cos(k) \\
\Gamma_2^{xz}(k) &= \phantom{-} 4 \Gamma_2^{xz} a(k)^2 \cos(2 k) \\
D_{2,a}^z(k) &= \phantom{-} 4 D_{2,a}^z a(k) a(k+\pi) \sin(2 k) \\
D_{2,a}^x(k) &= \phantom{-} 4 D_{2,a}^x a(k) a(k+\pi) \sin(2 k) \\
D_3^y(k, l) &= - 2 D_3^y a(k) a(k+\pi) \sin(2 \pi l).
\end{align}
\end{subequations}
Inspecting the above matrices one realizes that for zero magnetic field
 the slightly simpler form
\begin{equation}
\mathcal{M}_{k, l} = \begin{bmatrix}
A(k, l) & B(k, l) \\
B(k, l) & A(k, l)
\end{bmatrix}
\end{equation}
holds.\\ \ \\  \ \\  

\bigskip

{\noindent \textbf{\large Supplementary Note 3: Symplectic product and Berry 
phase for bosons}}\\

\noindent The Berry phase in quantum mechanics is defined by means of the
complex phase of the scalar product between two quantum states \cite{s_berry84}.
Thus, the key step is to define an appropriate scalar product.

In the main article, we use a symplectic product (9)
for the coefficients of bosonic creation and annihilation operators.
Note that this is a description on the level of second quantisation.
Here we want to elucidate more of its formal properties.
To be as general as possible, we consider a set of bosonic 
annihilation operators $a_j$ and creation operators $a^\dag_j$
with $j\in\{1,2,\ldots m\}$.
A general linear combination $c$ reads
\begin{equation}
\label{eq:gen-lin-combi}
c := \sum_{j=1}^m (u_j a_j^\dag - v_j a_j)
\end{equation}
where $c$ is not normalized and it is not specified whether it 
is a creation or annihilation operator.
Then, we define the corresponding generalized ket state by
the column vector
\be
\label{eq:gen-ket}
|c\rangle\rangle := (u_1,\ldots,u_m,v_q,\ldots,v_m)^\top =\underline{c}.
\ee
Sometimes the vector notation $\underline{c}$ is more convenient than the 
ket notation. Axiomatically, we can define the symplectic product
between two kets $|c\rangle\rangle$ and $|c'\rangle\rangle$ by
\bs
\label{eq:symplectic}
\begin{align}
\langle\langle c|c'\rangle\rangle
&:= \sum_{j=1}^m (u^*_j u'_j - v^*_j v'_j)
\\
&= \underline{c}^\dag  \eta \underline{c}',
\end{align}
\es
where the diagonal $2m\times 2m$ matrix 
$\eta=\text{diag}(1_1,\ldots,1_m,-1_{m+1},\ldots,-1_{2m})$
is used as a metric with $\eta^2=\mathds{1}$. 
This sort of ``generalized scalar product''
runs under several names in the literature such as
``quasi-scalar product'' or ``para-scalar product'' \cite{s_colpa78, s_kawag12, s_shind13}.
We prefer to avoid the term ``scalar product'' which suggests
semi-positivity, but use the established attribute ``symplectic''.
It is easy to verify that a conventional Hermitian matrix $M=M^\dag$ is not
self-adjoint with respect to Eqs.\ \eqref{eq:symplectic}. But 
$\eta M$ is self-adjoint due to
\bs
\begin{align}
\langle\langle c|\eta M c'\rangle\rangle &=
\underline{c}^\dag \eta\eta M\underline{c}' 
\\
&=\underline{c}^\dag  M\underline{c}' 
\\ 
\langle\langle \eta M c| c'\rangle\rangle &=
\underline{c}^\dag  M \eta\eta\underline{c}' 
\\
&=\underline{c}^\dag  M\underline{c}' .
\end{align}
\es

Alternatively, one can also start from
\be
\label{eq:equiv-symplect}
\langle\langle c|c'\rangle\rangle := [c^\dag,c']
\ee 
which obviously yields an expression identical to Eqs.\ \eqref{eq:symplectic}.
We observe that $\langle\langle c|c\rangle\rangle >0$ tells us
that $c$ is an unnormalised creation operator while 
$\langle\langle c|c\rangle\rangle <0$ tells us that it is 
an unnormalised annihilation operator.

The following question is imminent at this stage:
Can one relate Eqs.\ \eqref{eq:symplectic} and Eq.\ \eqref{eq:equiv-symplect}
to the conventional scalar product between quantum states?
The answer is ambiguous: it depends.
If there is a general ground state, i.e.\  a vacuum $\ket{0}$
annihilated by all annihilation operators $b$ considered (here the linear combinations $b$ have to be annihilation operators), then
the following relation between the standard scalar product 
$\bra{0} b' b^\dag \ket{0}$ in Fock space for two one-particle states
and the above defined symplectic product holds
\bs
\label{eq:berry-bz}
\begin{align}
\bra{0} b' b^\dag \ket{0} &= \bra{0} (b' b^\dag - b^\dag b') \ket{0}
\\
&=[b',b^\dag]
\end{align}
\es
where the last line is precisely definition \eqref{eq:equiv-symplect}
equivalent to \eqref{eq:symplectic}. Indeed, this situation is a very common
one in multi-band systems where $\ket{0}$ is the vacuum with respect
to all bosons at all values of $\mathbf{k}$. Then one retrieves
the Berry connection (10) and the Berry phase (11) for paths through
the Brillouin zone in the main text.

But we stress that the identity \eqref{eq:berry-bz} does not
hold if an external control parameter $\lambda$ is varied 
which changes the vacuum as well. Then the Berry phase for a 
path from $\lambda=0$ to $\lambda=\lambda_1$ reads 
\bs
\label{eq:berry-general}
\begin{align}
\Omega &= \mathbf{i} \int_{0}^{\lambda_1} 
\bra{0(\lambda)} b(\lambda) \partial_\lambda b^\dag(\lambda) \ket{0(\lambda)} 
d\lambda
\\
&= \mathbf{i} \int_{0}^{\lambda_1} 
\left[ 
\bra{0(\lambda)} b(\lambda) \left\{ \partial_\lambda b^\dag(\lambda)\right\}
 \ket{0(\lambda)} + \bra{0(\lambda)} b(\lambda)  b^\dag(\lambda) 
\left\{\partial_\lambda \ket{0(\lambda)}\right\} \right]d\lambda
\\
&= \mathbf{i} \int_{0}^{\lambda_1} 
\left[ 
\bra{0(\lambda)} [b(\lambda), \left\{\partial_\lambda b^\dag(\lambda)\right\}]
 \ket{0(\lambda)} + \bra{0(\lambda)} \left\{\partial_\lambda \ket{0(\lambda)}
\right\} \right] d\lambda
\\
&= \Omega_{\text{exc}}(\lambda_1) + \Omega_{\text{vac}}(\lambda_1)
\end{align}
\es
where two contributions are identified
\bs
\begin{align}
\Omega_{\text{exc}}(\lambda_1) &:= 
\mathbf{i} \int_{0}^{\lambda_1} 
\bra{0(\lambda)} [b(\lambda), \left\{\partial_\lambda b^\dag(\lambda)\right\}]
 \ket{0(\lambda)} d\lambda
\\
\Omega_{\text{vac}}(\lambda_1) &:= 
\mathbf{i} \int_{0}^{\lambda_1} 
 \bra{0(\lambda)} \left\{\partial_\lambda \ket{0(\lambda)}
\right\} d\lambda.
\end{align}
\es
One, $\Omega_\text{exc}$, 
results from the bosonic excitation
and equals what one obtains using the symplectic product.
The other, $\Omega_\text{vac}$, is the Berry phase of the vacuum.
For paths in the Brillouin zone the analogous result has been 
derived in Ref.\ \cite{s_peano16} where, however, the 
vacuum contribution should not occur because the global vacuum of the
system does not depend on momentum.

The bottom line is that for topological properties defined on the 
Brillouin zone the symplectic product yield a Berry phase identical
to the conventional definition. In more general cases, however,
the variation of the vacuum matters as well.

We corroborate this conclusion by repeating Berry's original adiabatic
approach in the bosonic Fock space. Let us assume that 
the bilinear Hamiltonian depends on the parameter $\lambda$ 
which may parametrises a path in the Brillouin zone or may be an
external control parameter. It is varied slowly from
$0$ to $1$, i.e.\ $\lambda=t/T$ for $t\in[0,T]$ with $T\to\infty$. 
The Hamiltonian is generally given by 
the matrix $\mathcal{M}(\lambda)$ \cite{s_blaiz86}; for an example
see Eq.\ (4) in the main article. At each value of $\lambda$
the ket $|n(\lambda)\rangle$ parametrises the creation of a boson
in the $n$th eigen mode. Hence the equation 
\be
\eta \mathcal{M}(\lambda) \ket{n(\lambda)}\rangle = 
\omega_n(\lambda) \ket{n(\lambda)}\rangle
\ee
is fulfilled. We assume the eigen modes to be non-degenerate for clarity.
The adiabatic ansatz, see for instance Ref.\
\cite{s_uhrig91}, for the solution $\ket{\psi_n(t)}$
 close to the instantaneous eigen state 
$\ket{\phi_n(\lambda)}:=b^\dag_n(\lambda)\ket{0(\lambda)}$ reads
\be
\ket{\psi_n(t)} = \exp(-i\Theta(t)) \left(\ket{\phi_n(\lambda(t))} + (1/T)
\ket{\perp}\right)
\ee
where the correction $(1/T) \ket{\perp}$ 
is small in $1/T$ and perpendicular to $\ket{\phi_n(\lambda(t))}$.
Inserting this ansatz in the Schr\"odinger
equation $\mathbf{i}\partial_t \ket{\psi_n(t)} = H \ket{\psi_n(t)}$
yields
\be
H\ket{\psi_n(t)} =(\partial_t \Theta)\ket{\psi_n(t)}
+ \exp(-i\Theta(t)) \frac{\mathbf{i}}{T}
\partial_\lambda \ket{\phi_n(\lambda)} + \text{perpendicular terms}.
\ee
Next, we multiply with $\bra{\phi_n(\lambda)}$ from the left to obtain
\be
\omega_n(\lambda) +E_0(\lambda) = \partial_t \Theta+ \frac{1}{T}
\partial_\lambda(\Omega_\text{exc}(\lambda)+\Omega_\text{vac}(\lambda))
\ee
where $E_0$ is the ground state energy and 
we used the result of the calculation \eqref{eq:berry-general}.
Integrating from $\Theta(t=0)=0$ to $t=T$ yields
\be
\Theta(T) = T \int_0^1 (\omega_n(\lambda) +E_0(\lambda))d\lambda
-\Omega_{exc}(1)-\Omega_{vac}(1).
\ee
This is the usual result for Berry phases in an adiabatic setting.
The first term represents the dynamic phase and the second term 
$\Omega_\text{exc}(1)+\Omega_\text{vac}(1)$ is the Berry phase. 
Clearly, there is a contribution from
the excitation and potentially from the ground state, i.e.\ the bosonic vacuum.
Again, if the ground state is a global vacuum applying to all bosons
 it does not change as a function of $\lambda$.
Then there is no vacuum Berry phase, i.e.\
$\Omega_\text{vac}=0$. 
This is the case for topological phases determined in the Brillouin zone.

\bigskip

{\noindent \textbf{\large Supplementary Note 4: Numerical calculation of the Zak phase}}\\

\noindent  
Only in rare cases, the analytical determination of the Zak phase is possible.
In particular for higher dimensional problems, for instance the twelve
dimensional extended model considered for \bcpo, a numerical approach 
is needed.  The first step is to discretise the contour of integration.
As an example for determining the phase from $k=0$ to $k=2\pi$ we
use $k_i = \frac{2 \pi i}{N}$ with $i=0, 1, \cdots, N-1$ (lattice constant is
set to unity).
It is straightforward to determine the eigen modes $\ket{n,k_i}\rangle$
numerically. But the numerical choice of phase at each momentum $k_i$
is arbitrary so that we cannot rely on a continuous evolution
and hence an approximation of
\be
\Omega_n = \mathbf{i} \int_0^{2\pi} \langle\bra{n,k} 
\partial_k \ket{n,k}\rangle dk
\ee
does not work. A well-established solution \cite{s_grusd14, s_wilcz84} consists in using the Wilson
loop 
\be
\label{eq:wilson-loop}
\Omega_n = - \mathrm{Im} \sum_{i=0}^{N-1} \, 
\ln\left(\langle \braket{n, k_i| n, k_{i+1}} \rangle \right) 
\quad \mathrm{mod} \; 2 \pi
\ee
instead, where $\ket{n,k_0}\rangle=\ket{n,k_N}\rangle$ holds because the loop
is closed. We stress that in the above formula the gauge, i.e.\ the choice
of the phase, of each eigen mode does not matter because it cancels.
Re-gauging each eigen mode arbitrarily
\be
\label{eq:re-gauge-arbitrary}
\ket{n,k_j}\rangle \to \widetilde{\ket{n,k_j}\rangle} = 
\exp(\mathbf{i}\varphi_j) \ket{n,k_j}\rangle
\ee
does not alter the outcome of Eq.\ \eqref{eq:wilson-loop} because each eigen mode
appears once as ket and once as bra.

An alternative variant of the above approach relies on the idea 
of parallel transport. The eigen mode $\ket{n, k_j} \rangle$ serves as
 reference state for $\ket{n, k_{j+1}} \rangle$. If their symplectic product 
reads
\be
\langle \braket{n, k_j|n, k_{j+1}} \rangle = z = |z| 
\exp(-\mathbf{i} \varphi_{j+1})
\ee
we re-gauge $\ket{n, k_{j+1}} \rangle$ such that it becomes as parallel
as possible to $\ket{n, k_j} \rangle$. Obviously, this is achieved by
\be
\label{eq:re-gauge}
\ket{n,k_{j+1}}\rangle \to \widetilde{\ket{n,k_{j+1}}\rangle} =
\exp(\mathbf{i}\varphi_{j+1})\ket{n,k_{j+1}}\rangle.
\ee
This procedure is iterated recursively from $j=0$ to $j=N-2$. 
The next and final step for $j=N-1$
yields $\varphi_N$, but the corresponding re-gauging \eqref{eq:re-gauge} is
not possible because the phase of $\ket{n,k_0}\rangle=\ket{n,k_N}\rangle$
is fixed already. Then the total sum \eqref{eq:wilson-loop}
simply reduces to 
\be 
\Omega_n = - \mathrm{Im} \ln\left(\langle \braket{n, k_{N-1}| n, k_{0}} 
\rangle \right) ,
\ee
since all re-gauged products are real and positive and the Zak phase corresponds to
\be
\Omega_n= \varphi_N.
\ee
The attractive feature of this second variant is that it reveals
the geometric character of the Berry phases. They stem from
the parallel transport in the U(1) principal fiber bundle of the 
manifold given by the eigen modes as functions of momenta.

\bigskip

{\noindent \textbf{\large Supplementary Note 5: Winding number $w$}}\\

\noindent In the case of the established minimal model for \bcpo{} 
the Hamiltonian shows an additional, chiral symmetry
which allows us to calculate the winding number even in
the presence of Bogoliubov terms. Here we show the details of 
the calculation of the winding number. 

In the minimal model with $D_3^y = 0$, the $12\times12$ matrix in Eq.\ (6) in the main article or 
in Eq.\ \eqref{eq:hamiltonian12} in Note 2
 can be split into $4 \times 4$ matrices 
simplifying the subsequent analysis which is performed
similarly to the one in Ref.\ \cite{s_joshi17}. To this end, we focus on
 the $x$-mode and its coupling to the $y_\pi$-mode. Since all couplings  
which are proportional to the $4 \times 4$ identity matrix do not alter the eigen 
modes they do not alter the topological properties and are therefore neglected. The 
coupling contributions proportional to $\sigma_x \otimes \mathds{1}$ only lead to 
small variations of the energy dispersion and we neglect them in a simplifying approximation. We  
checked that their omission has no impact on the Zak phase. We expect that the 
winding number similarly is not changed by the couplings proportional to 
$\sigma_x\otimes 1$. The same is assumed for the inclusion of small $D_3^y$. 
Thus, for simplicity, we
consider the Hamiltonian of single ladders
\begin{align}
\mathcal{H} = \frac{1}{2} \sum_{k} \mathbf{a}_{k}^\dagger \mathcal{M}_{k} 
\mathbf{a}_{k} 
\end{align}
with the Nambu spinor 
$\mathbf{a}_k = (t_{k}^x, t_{k+\pi}^y, t_{-k}^{x, \dagger}, 
t_{-k-\pi}^{y, \dagger})^{\top}$ and the  $4 \times 4$ matrix
\begin{equation}
\mathcal{M}_{k} = \begin{bmatrix}
C(k) & C(k) \\
C(k) & C(k)
\end{bmatrix} ,
\end{equation}
where the $2 \times 2$ matrix $C$ is parametrised by Pauli matrices
$\mathbf{\sigma}= \left( \sigma_x, \sigma_y, \sigma_z \right)$ 
\bs
\begin{align}
C(k) &= \mathbf{d}(k) \cdot \mathbf{\sigma} 
\\
\mathbf{d}(k) &= 
\left( 0, \Gamma_1^{xy}(k) - D_2^z(k), \frac{1}{2} \sum_{i=0}^2 (\Gamma_i^{xx}(k) - \Gamma_i^{yy}(k+\pi)) \right).
\end{align}
\es
Then, the chiral symmetry operator is easy to identify as 
$\mathds{1} \otimes \sigma_x$.  It fulfils the anticommutator 
$\left\lbrace \mathds{1} \otimes \sigma_x, \mathcal{M}_k \right\rbrace = 0$. In order to calculate the winding number we transform the Hamiltonian into the eigen basis of the chiral symmetry operator.
This is achieved by the unitary transformation
\begin{align}
U = \frac{1}{\sqrt{2}} \begin{pmatrix}
1 & 0 & 1 & 0 \\
1 & 0 & -1 & 0 \\
0 & 1 & 0 & 1 \\
0 & 1 & 0 & -1
\end{pmatrix}.
\end{align}
In this basis, the Hamiltonian matrix $\eta \mathcal{M}_k$ with 
the metric $\eta = \sigma_z \otimes \mathds{1}$ has a block off-diagonal form 
\bs
\begin{align}
\widetilde{\mathcal{M}}_k &= U^\dagger ( \eta \mathcal{M}_k ) U  
\\
&= \begin{bmatrix}
0 & D_1(k) \\
D^*_1(k) & 0
\end{bmatrix} .
\end{align}
\es
The  matrix $D_1(k)$ is given by
\be
D_1(k) = \begin{pmatrix}
d_3(k) + \mathbf{i} d_2(k) & d_3(k) + \mathbf{i} d_2(k)\\
- d_3(k) - \mathbf{i} d_2(k) & - d_3(k) - \mathbf{i} d_2(k)\\
\end{pmatrix}
\ee
and the winding number \cite{s_joshi15} is calculated by
\be
w = \frac{1}{8 \pi \mathbf{i}} \oint_\mathrm{BZ} \mathrm{d} k \mathrm{Tr}_2
\left( D^{-1} \partial_k D - (D^\dagger)^{-1} \partial_k D^\dagger \right)
\ee
with $D = \left( D_1(k) + D_1^\top(k) \right)/2$. 
By construction, the winding number is quantized to integer values 
$w\in\mathds{Z}$. For the investigated mode we find $w = -1$. 

The same analysis can be performed  for the $y$-mode coupled to the $x_\pi$-mode
yielding the same winding number. 
In contrast, the $z$-mode  only displays the trivial winding number 
$w=0$ because it does not couple with another mode.
Hence, it cannot be twisted or wound in any way.

A chiral symmetry of the general $12 \times 12$ matrix including all possible contributions could not be identified so that we could not define
a winding number in general. \ \\ \ \\ \ \\

\bigskip

{\noindent \textbf{\large Supplementary Note 6: (De)Localisation of the edge modes in the Su-Schrieffer-Heeger (SSH) model}}\\

\noindent In the main article, we provide three arguments why an edge state
generically delocalises if the system does not display an indirect gap
between the two bands where the eigen energy of the edge state
is located. Since the topology of bosonic systems  is still less
known and the model for \bcpo{} is rather intricate we want
to support our hypothesis on the delocalisation of edge states
by a transparent calculation for an established and well-known
fermionic model.

\begin{figure}
\centering
\includegraphics[width=\textwidth]{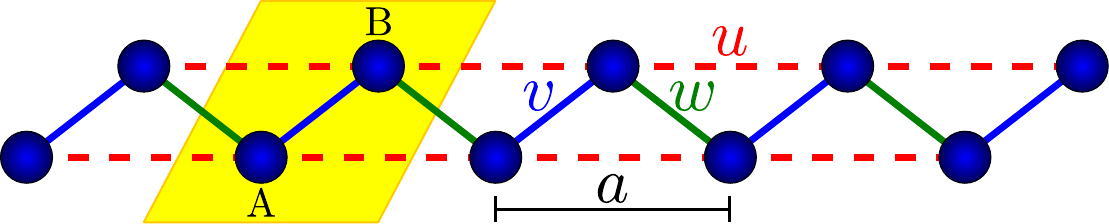} 
\caption{Lattice of the extended Su-Schrieffer-Heeger model with 
intracell coupling $v$, intercell couplings $w$ 
between nearest neighbour and intercell coupling $u$ between next-nearest neighbours. The unit cell comprising site $A$ and $B$ is displayed 
by the yellow area.}
\label{fig:ssh_lattice}
\end{figure}

\begin{figure}
\centering
\includegraphics[width=\textwidth]{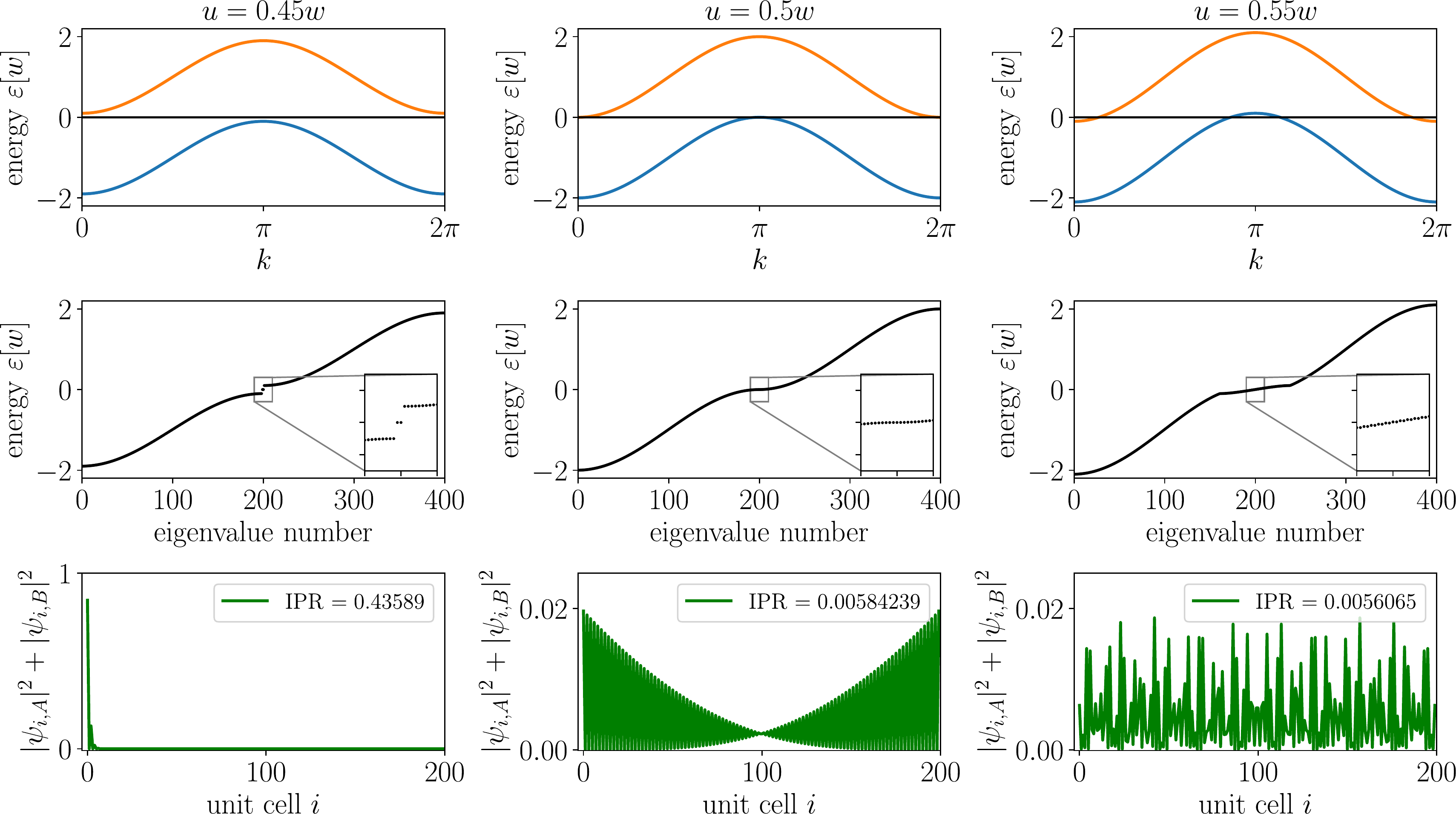} 
\caption{Delocalisation of the edge states illustrated for $v=0$. 
The three rows show from top
to bottom: the bulk dispersion, the eigen values of a finite piece of chain
with open boundaries consisting of 200 unit cells and the probability density 
$|\psi_{i,A}|^2+|\psi_{i,B}|^2$ of the eigen state $\psi$ 
with the highest inverse participation ratio (IPR) given in the legend
of the lower panels. The three columns refer to different ratios  $u/w$.
For $u=0.45 w$, the two edge modes lie
within the indirect gap and are well localised. The case $u=0.5w$ is
marginal and for $u=0.55w$ no indirect gap exists anymore. Concomitantly, 
no localised modes exist. But note that the two bands continue to be
clearly separated.}
\label{fig:ssh_ref}
\end{figure}

In order to do so we consider the SSH model \cite{s_su79} and extend it slightly 
by the coupling $u$ between next-nearest neighbours, see
Fig.\ \ref{fig:ssh_lattice}. Its Hamiltonian reads
\be
\mathcal{H}_\text{SSH} = \sum_i^N v \left(c_{i, B}^\dagger c_{i, A}^{\phantom{\dagger}} + 
w c_{i+1, A}^\dagger c_{i, B}^{\phantom{\dagger}} + u c_{i+1, A}^\dagger 
c_{i, A}^{\phantom{\dagger}} + 
u c_{i+1, B}^\dagger c_{i, B}^{\phantom{\dagger}}\right) + \text{h.c.} 
\ee
where $c_{i, A}$ is the fermionic annihilation operator on site $A$ of 
unit cell $i$ and $c_{i, B}$ the corresponding 
fermionic annihilation operator on site $B$.
The Hermitian conjugate operators are the creation operators. 
The three couplings are shown in Fig.\ \ref{fig:ssh_lattice}.
The Hamiltonian $\mathcal{H}_\text{SSH}$ is particle-conserving.
In the bulk or for periodic boundary conditions a Fourier transformation 
yields
\bs
\begin{align}
\mathcal{H} &= \sum_k \begin{pmatrix}
c_{k, A}^\dagger & c_{k, B}^\dagger
\end{pmatrix} \mathcal{M}_k
\begin{pmatrix}
c_{k, A}^{\phantom{\dagger}} \\
c_{k, B}^{\phantom{\dagger}}
\end{pmatrix} 
\\
\mathcal{M}_k &= \begin{pmatrix}
2 u \cos(k) & v + w \mathrm{e}^{\mathrm{i} k} \\
v + w \mathrm{e}^{-\mathrm{i} k} & 2 u \cos(k)
\end{pmatrix} ,
\end{align}
where the lattice constant $a$ is set to unity.
\es
The ensuing dispersion is
\begin{align}
\varepsilon_n(k) = 2 u \cos(k) \pm \sqrt{v^2 + w^2 + 2vw \cos(k)} 
\end{align}
with $n\in\{1,2\}$ corresponding to the $\pm$ sign in front of
the square root. The dispersion branches
are depicted in the upper row of Fig.\ \ref{fig:ssh_ref} for $v=0$
and the indicated ratios $u/w$.

On the one hand, the eigen states are the same as in the usual SSH model without next-nearest neighbour coupling $u$ since the additional coupling leads to a modification
of the matrix $\mathcal{M}_k$ proportional to the $2\times 2$ identity matrix
$2u\cos(k)\mathds{1}_2$. For this reason, we 
call the coupling $u$ isotropic. 
The induced  modification does not change the eigen states.
Hence, the extended SSH model shows the same Zak phase and the same
winding number as the non-extended SSH model. 

On the other hand, however, the numerical analysis of a finite piece of chain 
with open boundary condition reveals that the localisation of the
edge states is \emph{not}
protected against the isotropic coupling despite the fact that the direct gap does not close so that the two bands remain separated, see Fig.\ 
\ref{fig:ssh_ref}. By the naked eye one already discerns that the 
wave function with the largest value of the inverse participation ratio (IPR)
defined in Eq.\ (13) in the main article, see also Ref.\ \cite{s_krame93},
is localised if the energy of the edge modes lies well within
the indirect gap. But upon decreasing the indirect gap to zero 
for $u\to w/2$ the IPR drops to zero as well in the thermodynamic limit. Then 
it is obvious that the corresponding states are no longer localised. We emphasise 
that this does not contradict the argument of bulk-boundary correspondence which 
simply requires that the energy gap has to close at the boundary to
another phase with a different quantized topological invariant.

This important result specifies the meaning of the wide-spread
used bulk-boundary correspondence more precisely.

\end{document}